\documentstyle[aps,multicol,prl]{revtex}
\begin{document}
\draft
\begin{title}
{\bf Three-Dimensional Elastic Compatibility: Twinning in Martensites} 
\end{title}
\author{K.~\O. Rasmussen, T. Lookman, A. Saxena, A.~R. Bishop, and R.~C. Albers} 
\address{Theoretical Division,
Los Alamos National Laboratory, Los Alamos, New Mexico 87545}
\maketitle
\begin{abstract}
{We show how the St.Venant compatibility relations for
strain in three dimensions lead to twinning for the cubic to tetragonal
transition in martensitic
materials within a Ginzburg-Landau model in terms of the  
six components of the symmetric strain tensor.   
The compatibility constraints generate an anisotropic long-range 
interaction in the order parameter (deviatoric strain) components.
In contrast to two dimensions, the free energy is characterized by  
a ``landscape" of  competing metastable states. We find a variety of
textures, which result from the elastic frustration due to the effects of
compatibility.  Our results are also applicable to structural phase 
transitions in improper ferroelastics such as ferroelectrics and 
magnetoelastics, where strain acts as a secondary order parameter. }

\end{abstract}
\date{\today}
\pacs{81.30.Kf, 64.70.Kb, 61.72.Dc, 82.65.Dp}
\begin{multicols}{2}
Strain plays a crucial role in many structural phase transitions, either as 
a primary order parameter (OP), e.g., in technologically important martensites 
\cite{mart} and shape memory alloys \cite{shape}, or as a secondary OP, e.g., in 
ferroelectric \cite{ferro} and magnetoelastic \cite{magnet} materials.  Moreover, 
self-consistent coupling of strain to either charge (e.g., high-temperature 
superconductors \cite{htc}) or spin (e.g., colossal magnetoresistance manganites
\cite{cmr}) degrees of freedom determines key physical properties.  However, 
the physical degrees of freedom in a continuum elastic medium (such as described by 
Ginzburg-Landau models) are contained in the displacement field, $u(r)$, even 
though it is the strain fields, $\epsilon_{ij}$, which appear in the free energy.  
Instead of treating the strains as independent fields, one must assume that 
they correspond to a physical displacement field, i.e., they are derivatives 
of a single continuous function.  This is achieved by requiring that they 
satisfy a set of nontrivial (Saint-Venant) compatibility relations concisely 
expressed by the equation $\nabla\times (\nabla\times\epsilon)=0$ \cite{compat} 
(in the absence of voids, dislocations, etc.). Indeed, ignoring the geometrical 
compatibility constraint and minimizing the free energy directly would lead 
to the erroneous result that non-OP strain components are identically zero, 
and that the OP responds trivially to perturbations, such as stress or local disorder,
which is incorrect.

We show for the first time that a consistent Ginzburg-Landau (GL) 
understanding of martensitic textures in terms of the OP-strain alone, and 
involving long-range OP strain-strain forces, is possible in three 
dimensions (3D).  A 3D free energy expressed entirely in the 
order parameter strain variable(s), rather than the displacement field,
provides 
a unified understanding of martensitic textures. We use 3D compatibility 
equations, linking the strain tensor components in the bulk and at 
interfaces, that induce anisotropic order-parameter strain interactions 
({\it masked} in a conventional displacement-field treatment). These two
long-range bulk and interface potentials, together with local compositional
fluctuations, drive the formation of global elastic textures. In contrast
to 2D, the rich
microstructure is the result
of frustration and competing metastable states in a complicated energy
landscape.  Here we 
will focus only on twinning for illustrative purposes.  Other complex 
textures follow from our formalism and will be reported elsewhere. Our work is an 
essential step towards understanding results of
High Resolution Electron Microscopy (HREM) and
neutron scattering studies in alloys such as 
FePd and NiAl. In $Fe_{x}Pd_{1-x}$, (011) twins preferentially grow at the expense of (101) twins as
the temperature is gradually lowered to the martensitic transition temperature\cite{FePd}.
In $Ni_{x}Al_{1-x}$, martensite forms with a long period stacking
structure involving (110) planes\cite{NiAl}. 
Due to the directional aspects of the long-range interactions encoded by the 
underlying symmetry of the lattice, our analysis predicts such twinning planes and
directions.

{\it Model.} As a specific example, we examine a cubic
to tetragonal transition. In this case the symmetry-adapted strains $e_i$ 
can be defined in terms of the Lagrangian strain tensor components $\epsilon_{ij}$ as the 
dilatation $e_1=(1/\sqrt{3}) (\epsilon_{11}+\epsilon_{22}+\epsilon_{33})$, 
the deviatoric (i.e., OP) strains $e_2=(1/\sqrt{2})(\epsilon_{11} 
-\epsilon_{22})$, $e_3=(1/\sqrt{6}) (\epsilon_{11}+\epsilon_{22} 
-2\epsilon_{33})$ and the three shear strains $e_4=2\epsilon_{23}$, $e_5 
=2\epsilon_{13}$, $e_6=2\epsilon_{12}$. The elastic energy in 3D is given by\cite{bk} 
$$F=F_L(e_2,e_3)+F_{grad}(\nabla e_2,\nabla e_3)+F_{cs}(e_1,e_4,e_5,e_6),$$  
where the Landau free energy (Fig. 1), which provides the driving force for the transition, is   
$$F_L=A(e_2^2+e_3^2)+Be_3(e_3^2-3e_2^2)+C(e_2^2+e_3^2)^2 ,$$ 
and the Ginzburg (or gradient) contribution, which is responsible for the interfacial energies, to the free energy is 
\begin{eqnarray}
F_{grad}&=&g\left [(e_{2,x}^2+e_{2,y}^2)+\frac{1}{3}(e_{3,x}^2+e_{3,y}^2)  \right . \nonumber \\
&+&\left .\frac{2}{\sqrt{3}}(e_{2,x}e_{3,x}-e_{2,y}e_{3,y})
+\frac{4}{3}e_{3,z}^2\right ] \nonumber \\ 
&+&h\left [(e_{3,x}^2+e_{3,y}^2)+\frac{1}{3}(e_{2,x}^2+e_{2,y}^2)    \right .\nonumber \\
&-&\left .\frac{2}{\sqrt{3}}(e_{2,x}e_{3,x}-e_{2,y}e_{3,y}) +\frac{4}{3}e_{2,z}^2 \right ] \nonumber 
\end{eqnarray}
The harmonic elastic energy contribution due to non-OP compression-shear (CS)
strain components is 
$$F_{cs}=A_c e_1^2+A_s(e_4^2+e_5^2+e_6^2) .$$ 
It is also possible to add a coupling, $F_{compos}(\eta,\nabla\eta,e_2,e_3)$, of 
the OP strain to compositional fluctuations $\eta$, which 
are responsible for tweed, but we do not consider such effects in the present paper.
\begin{figure}[t]
\vspace*{270pt}
\hspace{-45pt}
\includegraphics{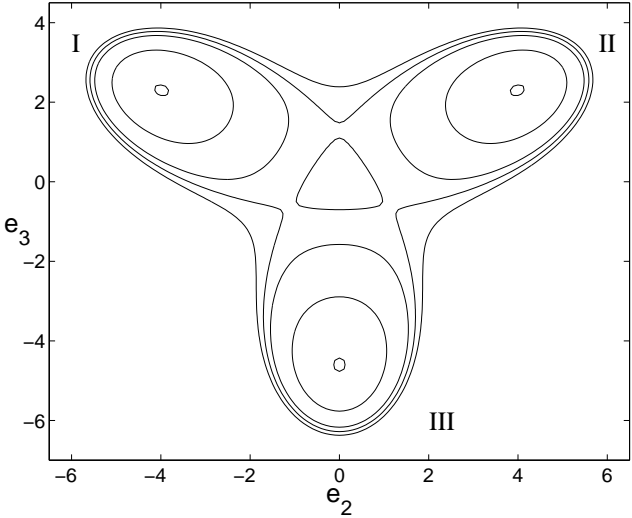}
\vspace*{-70pt}
\caption{Countour plot of the Landau free energy in the martensite phase
depicting the three degenerate energy minima corresponding to the three
tetragonal variants with tetragonal axis along the $x$, $y$ and $z$ axis,
respectively. Parameters are $A=-1.0$, $B=0.83$ and $C=0.04$.}
\label{fig1}
\end{figure}
The coefficients $A=A_o(T-T_o)$, $B$ and $C$ are related to the second,
third and fourth order 
elastic constants, respectively, and can be obtained from experimental 
structural data.  The martensitic transition temperature is denoted $T_o$ and 
the gradient coefficients $g$ and $h$ are determined from the 
phonon dispersion at the appropriate high-symmetry point in the 
Brillouin zone of the material.  $A_c$ and $A_s$ denote the bulk compression and shear 
moduli, respectively. Here we do not consider the 
gradient energy contribution from 
the non-OP components $e_1$, $e_4$, $e_5$, and $e_6$, since their contribution 
is of secondary nature compared to the terms reported above.

{\it 3D compatibility and analysis.} 
We express the non-OP strains in terms 
of the OP strains ($e_2$, $e_3$) using the three compatibility constraints  
$\epsilon_{12,12}=\epsilon_{11,22} + \epsilon_{22,11}$,
$\epsilon_{23,23}=\epsilon_{22,33} + \epsilon_{33,22}$,
$\epsilon_{31,31}=\epsilon_{33,11} + \epsilon_{11,33}$.
Analogous to the 2D case \cite{shenoy} the constrained  
minimization of the compression-shear energy
leads to an {\it anisotropic long-range interaction} between the OP 
strains  in the bulk:     
$$F_{cs}^{bulk}=\int d^3r\, e_{\alpha}(\alpha)U_{\alpha\alpha'}(|r-r'|) 
e_{\alpha'}(r')~~~(\alpha,\alpha'=2,3), $$ 
where the kernels $U_{22}$, $U_{23}$, and  $U_{33}$ can be expressed explicitly in Fourier space
with each of the three kernels becomes
a ratio of algebraic combinations of $k_x$, $k_y$ and $k_z$.
All these combinations have the same order in the length, $k$, of the wavevector 
$\vec{k}$. Hence, the kernels do not have any scale dependence. That is, the kernels
only depend on the angles $\phi$ and $\theta$, defined in terms of spherical 
coordinates.  The minima of each kernel occur at certain $\theta$ and $\phi$ values. These
angles vary from kernel to kernel and the system, in general, experiences
frustration. Figure 2 shows contours of the kernel $U_{22}$ and $U_{33}$
as functions of the two angles $\theta$ and $\phi$. 
The minima 
define a direction in 3D in which the compression-shear energy arising from
the  particular kernel is minimal.
 
In contrast to 2D, $F_{cs}$ can be viewed as a weighted sum of
the three kernels, where the fields $e_2$ and $e_3$ act as the weights. There is thus
an interplay between  $F_{L}$ and the directional dependence of $F_{cs}$ to accommodate
the frustration and choose minima, which in general do not have zero energy.
Thus, there exist
metastable minima corresponding to different microstructures.  In 2D
there is no
such interplay, as the directional minimum is independent of the amplitude
of the field:
There is only one minimum corresponding to zero energy and hence there is no
frustration. Moreover, these long-range kernels in 3D fulfill the full cubic
symmetry properties associated
with the transformation. In particular, $F_{cs}$ is invariant with respect to
two-fold, three-fold, and four-fold rotation symmetries of the cube. We do not present here the 
mathematical structure underpinning this,
but it is reflected in the frustration seen in 3D that is lacking in 2D.

The bulk contribution comes from a solid with periodic boundaries while the 
the effects of interfaces/surfaces such as habit planes (i.e., austenite-martensite or
parent-product boundaries) may be included
through a free energy contribution, $F_{cs}^{surf}$, where
$F_{cs}=F_{cs}^{bulk}+F_{cs}^{surf} $
and\cite{shenoy}
$$F_{cs}^{surf}=\int d^3k\, \Gamma(k_x,k_y,k_z)[e_2(k)^2+e_3(k)^2] .$$ 
In analogy with the 2D case the function $\Gamma(\vec{k})$ has the
form
$$\Gamma(\vec{k})=\nu\left[\frac{1}{|k_x\pm k_y|}+\frac{1}{|k_y\pm k_z|}
+\frac{1}{|k_z\pm k_x|}\right] $$
for the [110] family of habit planes.

We remark that the physics of twinning  is quite different in 2D and 3D.
In 2D both the gradient and the bulk compatibility terms vary as $\sim1/r^2$
and thus do not lead to a length scale.  However, the surface compatibility
term in 2D is essential for introducing a length scale. It varies as $\sim1/r$
which then competes with the combined bulk
compatibility and gradient terms to give rise to a twinning width.
In 3D,  the gradient terms behave as $\sim 1/r^2$ and $F_{cs}^{bulk}\sim 1/r^3$.
Thus, there is already a length scale present and
$F_{cs}^{surf}$ merely renormalizes the length scale arising from the 
bulk compatibility and gradient terms. We have numerically verified the 
above scaling relations. On the basics of the above arguments we expect the twin width 
to scale as the square-root of the length in 3D to hold on the basis of
above arguments.

\begin{figure}[t]
\vspace*{245pt}
\hspace{-25pt}
\includegraphics{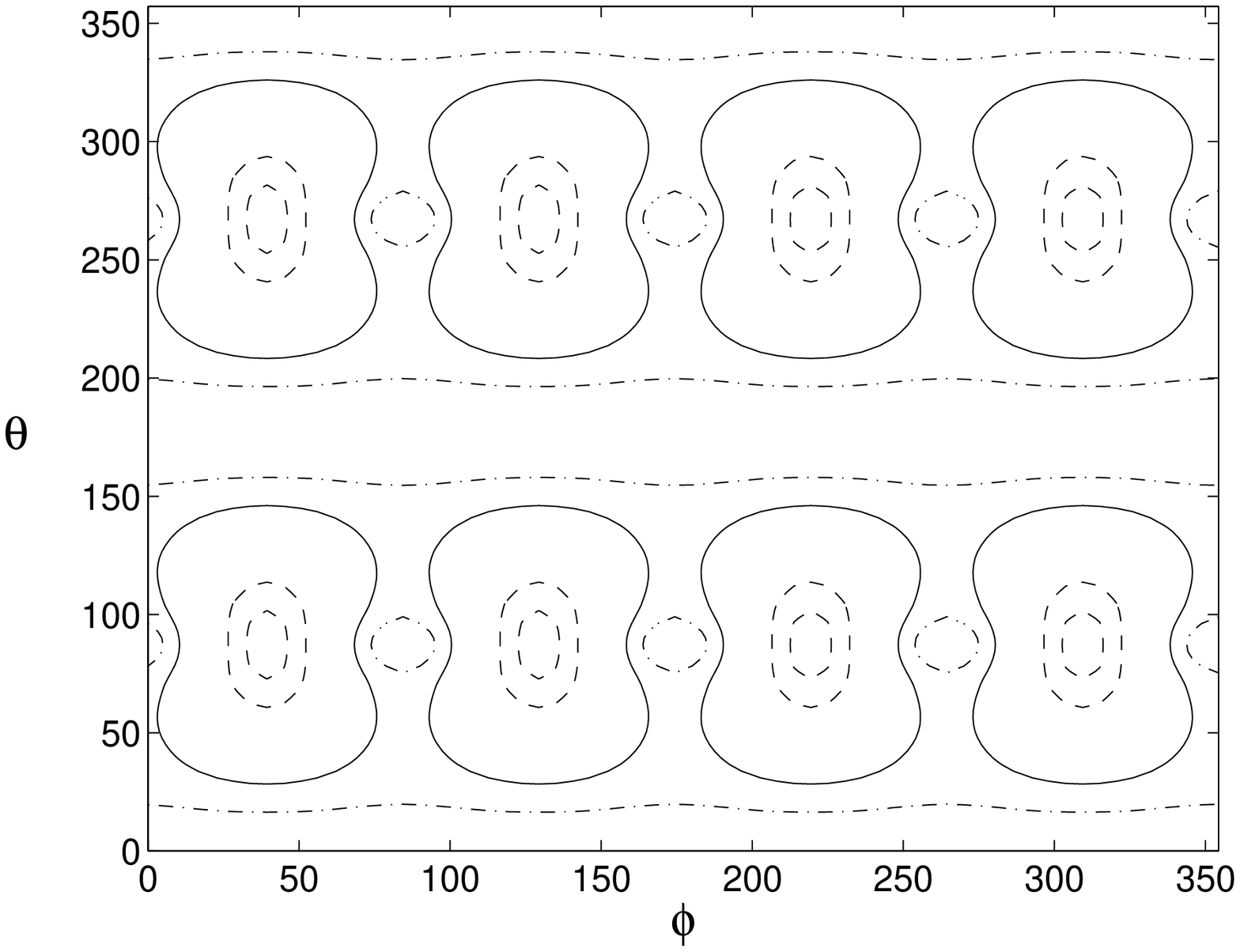}

\vspace*{180pt}

\hspace{-25pt}
\includegraphics{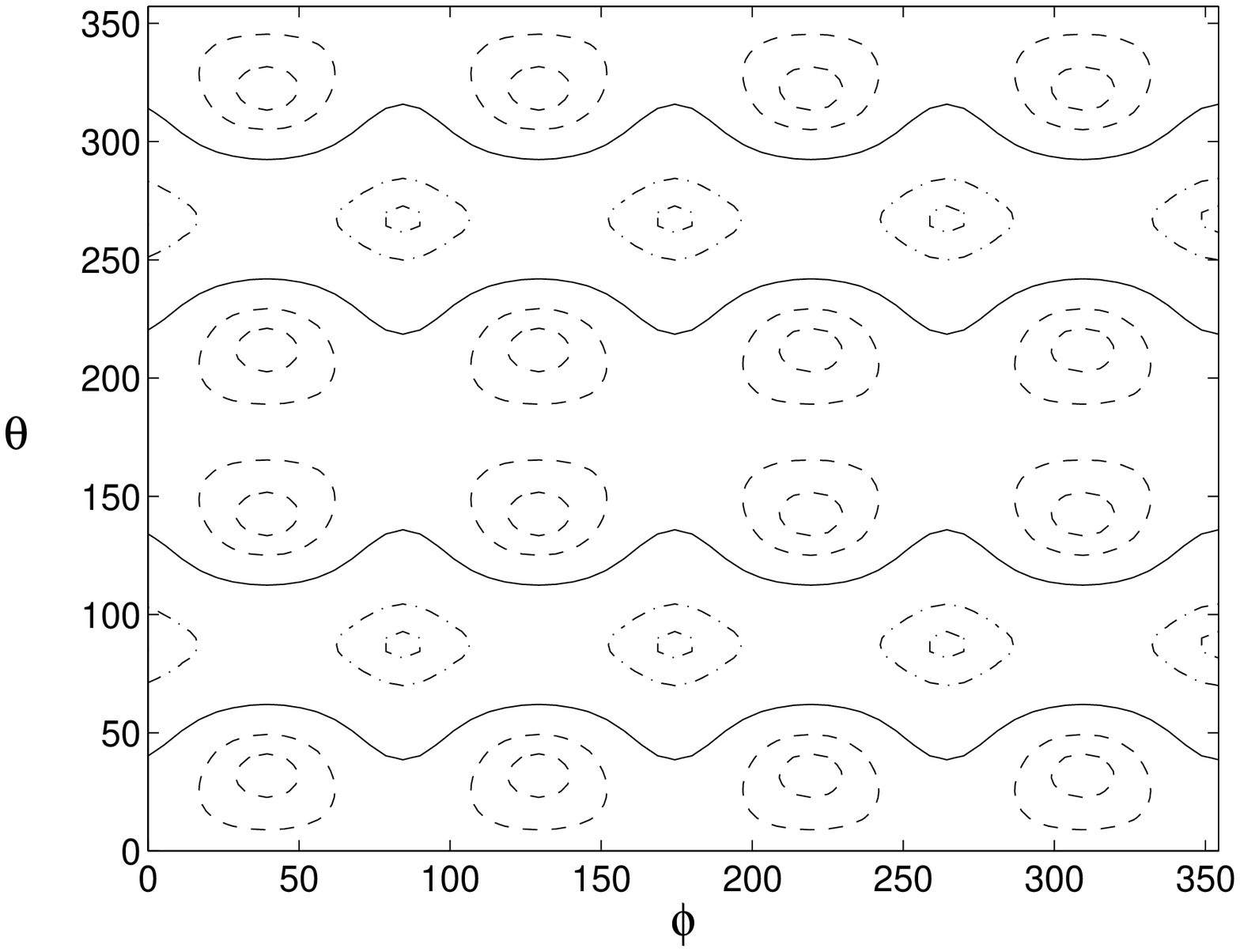}
\vspace*{-70pt}
\caption {Three dimensional compatibility kernels $U_{22}$ (upper) and $U_{33}$ (lower) as a function of the
colatitude $\theta$ and the azimuth $\phi$ angles. Dashed  contours surround minima, and dashed dotted surround
maxima, while solid contours show intermediate levels. The parameters are $A_s=1.2$ and $A_c=2.4$. }
\label{fig2}
\end{figure}

   
{\it Texture simulations.}
The various martensitic textures that realize minima of the energy are found  from 
random initial conditions and relaxational dynamics for the OP strains. That is, 
$
\dot{e}_{\alpha}(r)=-\frac{\delta F(e_{\alpha},e_1(e_{\alpha}),
e_4(e_{\alpha}),e_5(e_{\alpha}),e_6(e_{\alpha}))}{\delta 
e_{\alpha}(r)},
$ 
where the time $t$ is scaled with a characteristic relaxation rate and $\alpha=2,3$. Depending on the details 
of the initial configurations, various twinning textures reflecting the metastability emerge as $\dot {e}_{\alpha}$ vanishes. 
The simulations are performed for  a regular cube, and 
periodic boundary conditions are applied in all six directions  
to obtain fully relaxed textures in $e_2(r)$ and 
$e_3(r)$.  
\begin{figure}[t]
\vspace*{280pt}
\hspace{-45pt}
\includegraphics{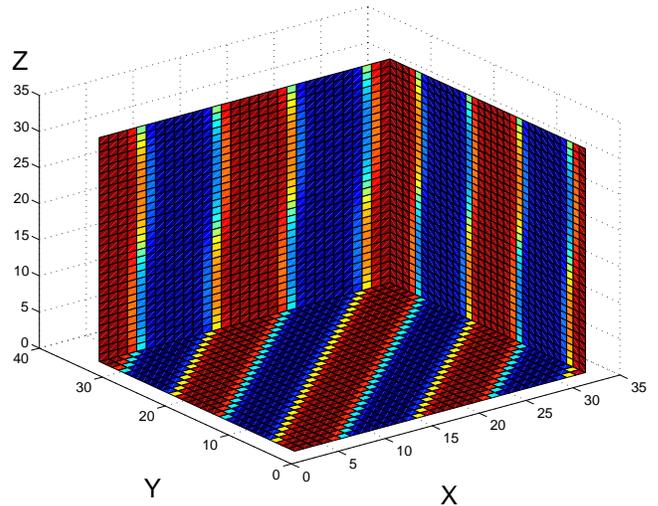}
\vspace*{-70pt}
\caption {(Color) Three dimensional twins in the (110) plane obtained from the 
time-dependent Ginzburg-Landau simulations, with representative parameters as given in Figs. 1 and 
2 and, $h=1$ and $g=1$. The scale of the twins is determined by the competition between the 
interfacial energy and the bulk compression-shear energy, whereas the sharpness
of the domain walls is determined by the Landau energy.} 
\label{fig3}
\end{figure}

We illustrate the simulation results obtained with representative 
(scaled) parameters.  The red/blue/green 
color represent ($e_2,e_3$) positive/negative/zero OP strain
values, respectively (the actual values 
are approximately determined by the minima of $F_L$ as 
illustrated in Fig.1).
   
Figure 3 shows the simplest twinning texture ((110) twins) that is uniform 
in the $z$ direction.  The strain $e_2$ changes from negative (minimum I in Fig. 1) 
to positive (minimum II in Fig. 1) (i.e. blue to red). The strain $e_3$ is 
not shown, as it is essentially constant with a small modulation at the twin boundaries.
This particular martensitic texture is the trivial extension of the 2D results since 
the OP fields are uniform in the $z$ directions. Notice also that the directional dependence 
is the same as observed in 2D studies \cite{shenoy}.

A richer texture is shown in Fig. 4,
where the  strain $e_3$ changes from negative (minimum III in Fig. 1) to positive 
(minimum I or II in Fig. 1) (i.e. blue to red). The strain $e_2$ would either be 
zero (green) when $e_3$ is negative (minimum III) or positive/negative when $e_3$ is 
positive.  The figure clearly shows that  twins with strain $e_2$ can be oriented
in two possible ways. This reflects the inversion symmetry of the kernels with  
respect to $x$, $y$.  We note that a similar
alternating twin microstructure has been implied by neutron and x-ray
scattering studies in layered high-temperature superconductors \cite{htc}.
\begin{figure}[h]
\vspace*{250pt}
\hspace{-25pt}
\includegraphics{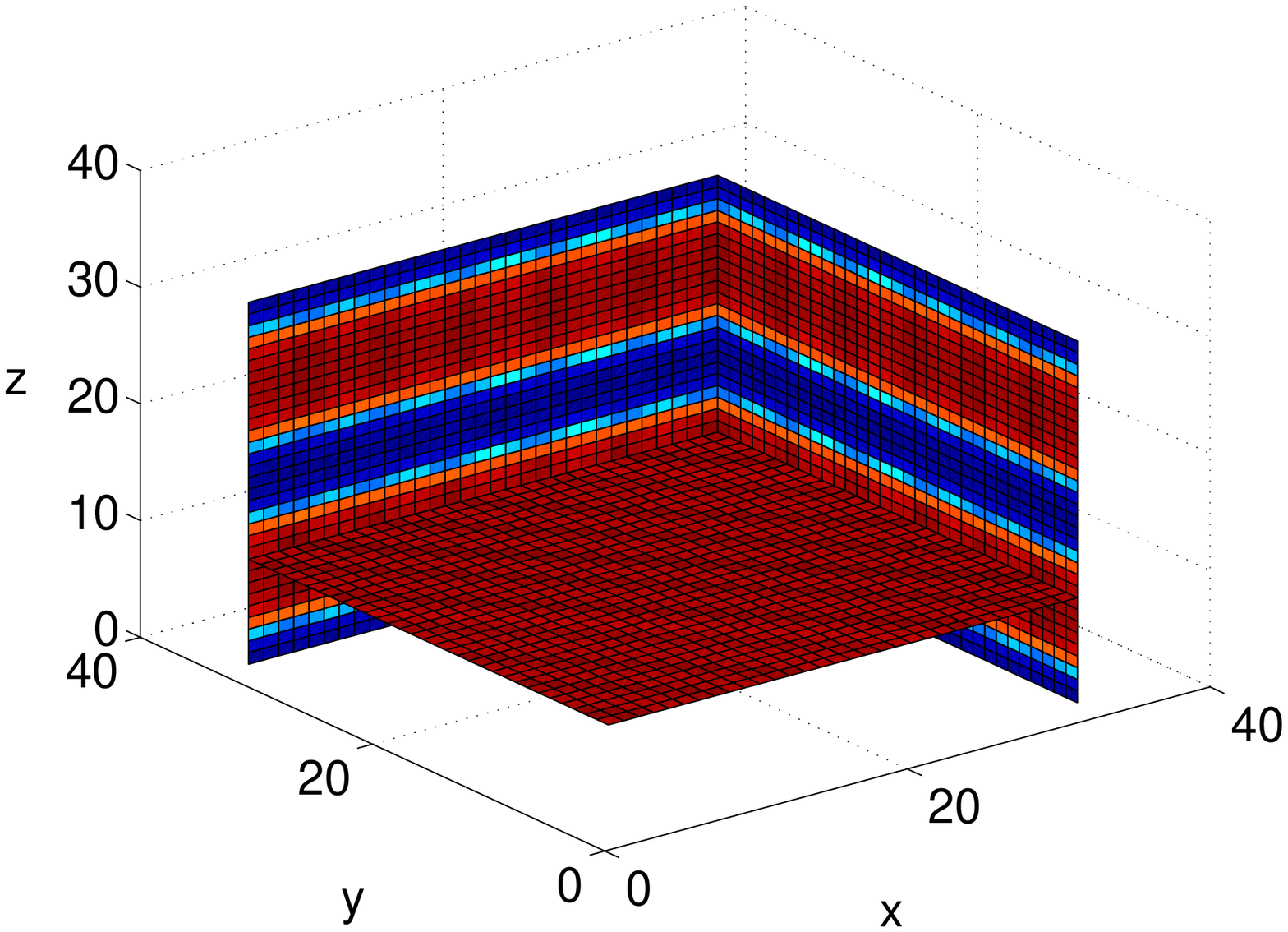}

\vspace*{170pt}

\hspace{-25pt}
\includegraphics{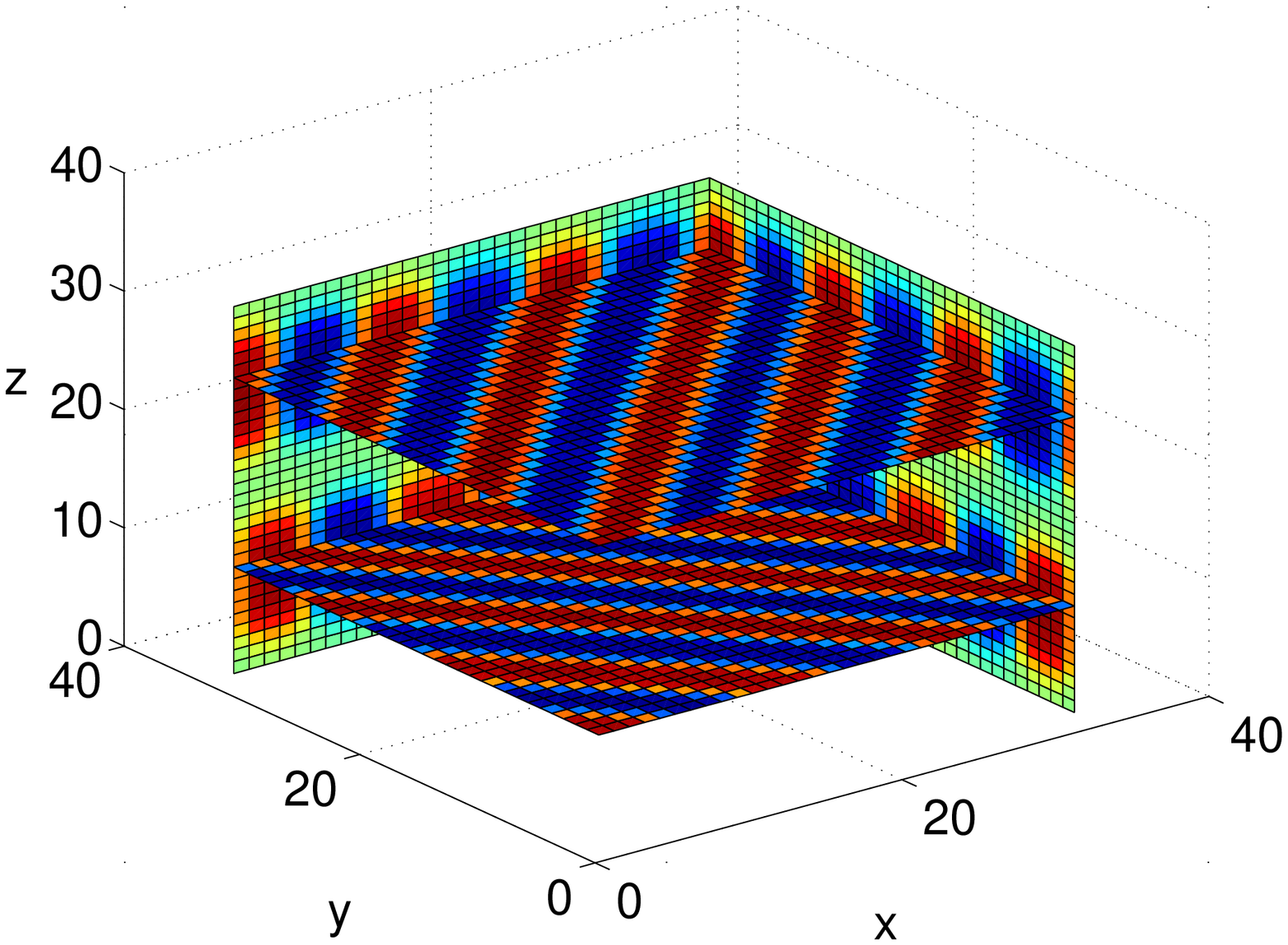}
\vspace*{-70pt}
\caption{(Color) Three dimensional twins with two different orientations, (110) 
and (1$\overline{1}$0), in adjacent planes. Both OP fields are shown, $e_3$ in the upper panel and 
$e_2$ in the lower.}
\label{fig4}
\end{figure}

To illustrate the of frustration that is inherent in the compatibility 
relations, we explain the orientation of the twins in Fig. 4.  We construct an 
effective kernel for the red and blue twins, respectively, using the appropriate 
values of the strains $e_2$ and $e_3$ in the twins.  Figure 5 depicts the minima 
of these effective kernels in the $\phi$-$\theta$ plane. The dashed line shows the 
minimum corresponding to the blue twins and the solid line the minimum for the red twins.  
We  note that the minima correspond to different $\phi$ angles ($45+\delta$ 
and $45-\delta$).  The shift $\delta$ depends on the parameters of the Landau 
free energy. Ideally, the red and the blue twins would 
not run parallel; however, the system accommodates these two conflicting directions  by choosing the 
average angle of 45 degrees.  This aspect of competing metastable states
is a novel feature associated with the cubic symmetry kernels that
leads to a rich landscape of microstructures.  
\begin{figure}[t]
\vspace*{250pt}
\hspace{-35pt}
\includegraphics{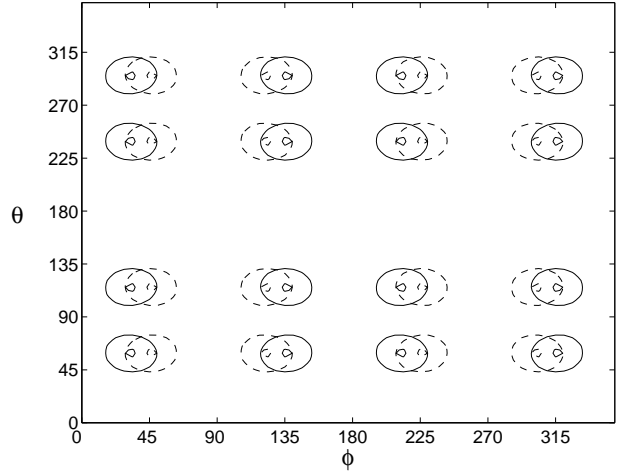}
\vspace*{-70pt}
\caption {Minima of effective kernels illustrating inherent strain 
(orientational) frustration in 3D.} 
\label{fig5}
\end{figure}

{\it Conclusion.}  We have derived a complete symmetry based, fully 3D model 
that describes the cubic to tetragonal structural transitions observed 
in many martensitic and shape-memory alloys.  We obtained analytically the 
compatibility-induced anisotropic long-range potential in the OP (deviatoric 
strain tensor components) in the bulk and at interfaces.  
Unlike the 2D case \cite{bhk,shenoy}, there is a length scale in the system  
from bulk compatibility alone; the surface compatibility potential 
also introduces a length scale akin to 2D.  Many twin orientations are possible in 
3D as a result of elastic frustration and the
``landscape" (probably ``rugged") of metastable energy states. Our model 
can be generalized straightforwardly to improper ferroelastics transitions  
by including symmetry allowed polarization (magnetization, etc.) nonlinear 
terms, and couplings to strain. Other symmetries can be handled within our 
approach.  For example, we can study the cubic to 
trigonal (rhombohedral) transition in lead orthovanade, and NiTi- and 
AuCd-based shape-memory alloys using the three shear strains $e_4$, $e_5$, and 
$e_6$ as OP and with $e_1$, $e_2$, and $e_3$ expressed in terms of the 
shear strains via compatibility. Image reconstruction methods
of systematic experimental data from HREM and neutron scattering will allow a more
quantitative comparison of
strain patterns with those obtained from our model.
  
We wish to thank G. R. Barsch, I. Mitkov, and S. R. Shenoy for 
insightful discussions. 
This work was supported by the U.S. Department of Energy (Contract number 
W-7405-Eng-36). This research used resources of the National Energy Research
Scientific Computing Center, which is supported by the Office of
Science of the U.S. Department of Energy under Contract
No. DE-AC03-76SF00098.

\end{multicols}
\end{document}